\documentclass[aps,preprint,prd,nofootinbib]{revtex4}
\usepackage{graphicx}
\usepackage{psfrag}

\begin{document}

\title{Two-Body Strong Decays of the $2P$ and $3P$ Charmonium states}
\author{
Zhi-Hui Wang$^{[1],[2],[3]}$\footnote{zhwang@nmu.edu.cn}, Guo-Li Wang$^{[4],[5]}$\footnote{ wgl@hbu.edu.cn}\\
}
\address{
$^1$ Key Laboratory of Physics and Photoelectric Information Functional Materials, North Minzu University, Yinchuan, 750021, China,\\
$^2$School of Electrical and Information Engineering, North Minzu University, Yinchuan, 750021, China\\
$^3$School of Physics and Center of High Energy Physics, Peking University, Beijing 100871, China\\
$^4$Department of Physics, Hebei University, Baoding, 071002, China,\\
$^5$Hebei Key Laboratory of High-precision Computation and
Application of Quantum Field Theory, Baoding, 071002, China
}

 \baselineskip=20pt

\begin{abstract}
Two-body open charm strong decays of
the $2P$ and $3P$ charmonium states are studied by the Bethe-Salpeter(BS) method combined with the $^3P_0$ model.
The wave functions and mass spectra of the $2P$ and $3P$ charmonium states are obtained
by solving the BS equation with the relativistic correction.
The strong decay widths and relative ratios of the $2P$ and $3P$ charmonium states are calculated.
Comparing our results with the experimental data,
we obtain some interesting results.
Considering the $X^*(3860)$ as the $\chi_{c0}(2P)$,
the total strong decay width is smaller than the experimental data.
But the strong decay width depends on the parameter $\gamma$ in the $^3P_0$ model,
and the mass and width of the $X^*(3860)$ have large errors,
we cannot rule out the possibility that the $X^*(3860)$ is the $\chi_{c0}(2P)$.
The $X(4160)$ is a good candidate for the $\chi_{c0}(3P)$,
not only the strong decay width of the $\chi_{c0}(3P)$ is same as the experimental data,
but the relative ratios $\frac{\Gamma(\chi_{c0}(3P)\to D\bar D)}{\Gamma(\chi_{c0}(3P)\to D^*\bar D^*)}\approx0.0019<0.09$,
and $\frac{\Gamma({\chi_{c0}(3P)\to D\bar D^*})}{\Gamma({\chi_{c0}(3P)\to D^*\bar D^*})}=0<0.22$
are consistent with the experimental results of the $X(4160)$.
Taking the $X(4274)$ as the $\chi_{c1}(3P)$,
the strong decay width is consistent with the experimental data,
so the $X(4274)$ is a good candidate for the $\chi_{c1}(3P)$.
Assigning the $X(4350)$ as the $\chi_{c2}(3P)$,
the corresponding strong decay width is slightly larger than the experimental data.
To identify if the $X(4350)$ is $\chi_{c2}(3P)$,
many more investigations are needed.
All of the strong decay widths and relative ratios of the $2P$ and $3P$ charmonium states can provide the
useful information to discover and confirm these particles in the future.
 \vspace*{0.5cm}

\end{abstract}

\maketitle
\section{Introduction}
Since the Belle Collaboration reported the first observation of the $X(3872)$~\cite{3872},
many more charmonium-like states have been observed experimentally.
Belle observed the $X(4160)$ from
the process $e^+e^-\to J/\psi D^*\bar D^*$,
which has the mass and width $M=(4156^{+25}_{-20}\pm15)$ MeV and $\Gamma=(139^{+111}_{-61}\pm21)$ MeV, respectively~\cite{39404160}.
They also gave the upper limits of relative ratios:
$\frac{B_{D\bar D}(X(4160))}{B_{D^*\bar D^*}(X(4160))}<0.09,$
$\frac{B_{D^*\bar D}(X(4160))}{B_{D^*\bar D^*}(X(4160))}<0.22$.
The $X(4140)$ was first observed by the CDF Collaboration in the exclusive decay $B\to J/\psi \phi K$~\cite{4140},
then another charmonium-like states the $X(4274)$ also was observed in the same decay channel~\cite{4274}.
These two charmonium-like states also were observed by LHCb Collaboration~\cite{lhcb1,lhcb2},
the mass and natural width of $X(4140)$ and $X(4274)$ were $M=(4146.8\pm2.4)$ MeV, $\Gamma=(22^{+8}_{-7})$ MeV,
and $M=(4274^{+8}_{-6})$ MeV, $\Gamma=(49\pm12)$ MeV~\cite{PDG}, respectively.
In 2010, BABAR observed the $Z(3930)$
in the $\gamma\gamma$ production of $D\bar D$ system~\cite{Z3930-BABAR}.
Now the Particle Data Group(PDG) gives the mass and
width of the $Z(3930)$ as $M=(3927.2\pm2.6)$ MeV and $\Gamma=(24\pm6)$ MeV~\cite{PDG}.
And the properties of $Z(3930)$ are consistent with the expectations for the
$\chi_{c2}(2P)$ state~\cite{chic2,slzhu1}.
Belle also explored a charmonium-like state $X(4350)$ in the process $J/ \psi \phi$ in 2010,
the extracted mass and width were $(4350.6^{+4.6}_{-5.1})$ MeV and $(13^{+18}_{-9}\pm4)$ MeV~\cite{4350}.
The $X^*(3860)$ was observed in the process $e^+e^-\to J/\psi D\bar D$ by Belle in 2017,
the corresponding mass and width are $M=(3862^{+26}_{-32}$$^{+40}_{-13})$ MeV and $\Gamma=(201^{+154}_{-67}$$^{+88}_{-82})$ MeV~\cite{3860},
respectively.

The properties of these charmonium-like states have inspired great interest in
both theoretical and experimental research fields of hadronic physics.
Many theoretical approaches have
studied the properties of these charmonium-like states~\cite{th1,th2,th201,th3,th4,th7,liux1,th9,th10,th12,th13,zhaoqiang,th6,Godfrey,liuxiang2021,Swanson}.
The Ref.~\cite{th201} investigated that the $X(4140)$ and $X(4274)$ can be both interpreted as the $S$-wave $cs\bar c\bar s$
tetraquark states of $J^P=1^+$.
The Ref.~\cite{th3} computed the
open-charm strong decay widths of the $\chi_c(3P)$ states and
their radiative transitions, and they suggested the $X(4274)$ could be interpreted as the $\chi_{c1}(3P)$ state.
Taking the $Z(3930)$ as $\chi_{c2}^\prime(2P)$,
the Ref.~\cite{liux1} investigated the decay $Z(3930)$ into $J/ \psi\omega$.
Considering the $X(4350)$ as the $\chi_{c2}^{\prime\prime}$, the Ref.~\cite{th13} analyzed the mass
and calculated the open charm strong decay of the $X(4350)$, which were
consistent with the existing experimental data.
The results of Ref.~\cite{th6} preferred the $J^{PC}=0^{++}$
assignment for the $X^*(3860)$ over the $2^{++}$ assignment,
which was also in agreement with the experiment.
Calculating the observable quantity (such as the spectrum or the strong decay width) of these charmonium-like states,
then comparing with the experimental data,
may help us to better understand the quark structure of the charmonium-like states.

In addition to the mass spectrum and strong decay width,
the electromagnetic decay also can help us to determine the structure these charmonium-like states.
According to E1 transition widths for the $\chi_{c1}(2P)\to\gamma J/\psi$ and $\chi_{c1}(2P)\to\gamma\psi(2S)$ and other results,
the Ref.~\cite{th8} argued that the $X(3872)$ may be a $\chi_{c1}(2P)$ dominated charmonium state with some admixture of the $D^0\bar D^{*0}$ component.
The Ref.~\cite{electromagnetic1} calculated the one- and two-photon decay widths of
$Y(3940)$, $Z(3930)$, $X(3915)$ and $X(4160)$ mesons.
Considering $X(4660)$, $X(3872)$, $X(3900)$, $X(3915)$ and $X(4274)$ as $5^3S_1$, $2^3P_1$,
$2^1P_1$, $2^3P_0$ and $3^3P_1$, respectively,
the Ref.~\cite{electromagnetic2} studied the E1 and M1 transition width, and annihilation decays of these charmonium states.
But the electromagnetic decay widths of these charmonium states are about the order of keV,
which are smaller than the results of Okubo-Zweig-Iizuka (OZI)-allowed strong decay.
These electromagnetic decays can only be detected experimentally when large amounts of data are available in the future. 
For now,
the strong decay widths and the relative ratios of these charmonium states are good ways to determine their properties.

In this paper we will focus on the strong decay widths of the $2P$ and $3P$ charmonium states.
The relativistic correction of the $2P$ and $3P$ charmonium states are larger than that of
the corresponding $1P$ states, therefore, we need a relativistic model.
The BS method is a relativistic framework that describes the bound state with definite quantum number,
the corresponding relativistic form of wave functions are the solutions of the full Salpeter equations.
Using the BS method,
we have discussed the properties of some radial excited states in previous work,
such as the semileptonic and nonleptonic $B_c$ decays to the $Z(3930)$ and $X(4160)$ as the $\chi_{c2}(2P)$ and $\chi_{c2}(3P)$~\cite{bcweak1},
the strong decays of the $X(3940)$ and $X(4160)$
as radial high excited states the $\eta_c(3S)$ and $\eta_c(4S)$~\cite{4S4160},
the radiative E1 decay of the $X(3872)$~\cite{thwang1,thwang2},
two-body strong decay of the $Z(3930)$ which was the $\chi_{c2}(2P)$ state~\cite{thwang}.
All the theoretical results are consistent with
the experimental data or other theoretical results.
So the BS method is a good way to describe the
properties and decays of the radially higher excited states.
In this paper,
we will study the strong decays of the $2P$ and $3P$ charmonium states by the BS method with the $^3P_0$ model.

For the $2P$ and $3P$ charmonium states,
the dominant decay is the Okubo-Zweig-Iizuka (OZI)-allowed two-body open charm strong decay.
We will adopt the $^3P_0$ model to calculate the two-body open charm strong decay.
The $^3P_0$ model was used to calculate the decay rates of the meson resonances in Ref.~\cite{3p01},
which assumed that the $q\bar q$ pair is produced from vacuum with quantum number $J^{PC}=0^{++}(^3P_0)$,
was applied to calculate the strong decay of heavy-light mesons~\cite{th13,3p0hl1}
and heavy quarkonia~\cite{Godfrey,3p05}.
We also studied the strong decays of some heavy quarkonia by the $^3P_0$ model combine with the BS method in Refs.~\cite{4S4160,thwang,fu},
the results were in accordance with the experimental data or other theoretical results.
So we take the same model to calculate the two-body open charm decay of the $2P$ and $3P$ charmonium states.

The paper is organized as follows.
We give the formulation of two-body strong decay of charmonium state in Section~II;
In Sec.~III, we show the numerical results and discussions;
We give the corresponding conclusions in Sec.~IV.
Finally, we present the instantaneous BS equation and the relativistic wave functions of $P$-wave charmonium states in the Appendix.

\section{two-Body Strong Decay of charmonium state}

\begin{figure}[htbp]
\centering
\includegraphics[height=3.5cm]{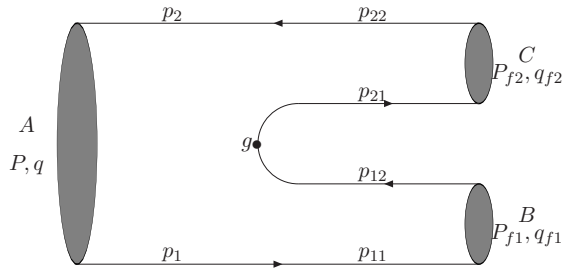}
\caption{\label{OZIStrongdecay}The Feynman diagram of two-body open charm strong decay.}
\end{figure}

To calculate the two-body open charm strong decays of the $2P$ and $3P$ charmonium states by the relativistic BS method,
we extend the $^3P_0$ model to relativistic form: $H=-ig\int d^4x\bar\psi\psi$~\cite{thwang,fu}.
Here $\psi$ is the dirac quark field, $g=2\gamma m_q$,
$m_q$ is the quark mass of the light quark-pairs,
$\gamma$ is a dimensionless constant that describes the pair-production strength.
In this paper, we take $\gamma=0.35$, which is the best-fit value for the usual $^3P_0$ model~\cite{Swanson}.
Combining the $^3P_0$ model with the BS wave functions of the initial and final mesons,
the corresponding amplitude in Fig.~\ref{OZIStrongdecay} can be written as
\begin{equation}\label{amp1}
\mathcal M=\langle BC|H|A\rangle=-ig\int\frac{{\rm d}^4q}{(2\pi)^4}{\rm Tr}
[\chi_P(q)S_2^{-1}(p_2)\bar\chi_{P_{f2}}(q_{f2})\bar\chi_{P_{f1}}(q_{f1})S_1^{-1}(p_1)]
\end{equation}

Integrating out the momentum $q_0$ with instantaneous approximation,
and neglecting all the negative energy contributions which have very small influence on the amplitude~\cite{fu},
then the leading order amplitude Eq.(\ref{amp1}) is the overlap integration of the positive BS wave functions for the initial and final states,
\begin{equation}\label{amp2}
\mathcal M=\langle BC|H|A\rangle=g\int\frac{{\rm d}^3\vec q}{(2\pi)^3}{\rm Tr}
\left [\frac{\not\!{P}}{M}\varphi_{p}^{++}(\vec q)\frac{\not\!{P}}{M}\bar\varphi_{p_{f2}}^{++}(\vec q_{f2})
\bar\varphi_{p_{f1}}^{++}(\vec q_{f1})\right ]\left(1-\frac{M-\omega_1-\omega_2}{2\omega_{12}}\right),
\end{equation}
where $\varphi^{++}_P(\vec q)$, $\varphi^{++}_{P_{f1}}(\vec q_{f1})$ and $\varphi^{++}_{P_{f2}}(\vec q_{f2})$
are the positive BS wave functions of the initial meson $A$, finial meson $B$ and $C$, respectively.
$\bar\varphi=\gamma^0\varphi^\dagger\gamma^0$.
$\vec q$, $\vec q_{f1}$, $\vec q_{f2}$
are the three dimensions relative momentum between the quark and anti-quark of the initial meson $A$,
finial meson $B$ and $C$, respectively.
$\vec{q}_{f1}=\vec q-\frac{m_c}{m_c+m_{u,d,s}}\vec{P}_{f1}$,
$\vec{q}_{f2}=\vec q+\frac{m_c}{m_c+m_{u,d,s}}\vec{P}_{f2}$.
$\vec{P}_{f1}$ and $\vec{P}_{f2}$ are the three momentum of finial mesons $B$ and $C$.
$w_{12}=\sqrt{m_{u,d,s}^2+\vec q^2_{f1}}$.

Finally, the two-body open charm strong
decay width of the $2P$ and $3P$ charmonium states can be expressed as
\begin{eqnarray}\label{width}
\Gamma=\frac{|\vec P_{f1}|}{8\pi M^2(2J+1)}\sum_{\lambda}|\mathcal M|^2,
\end{eqnarray}
where $|\vec P_{f1}|=\sqrt{[M^2-(M_{f1}-M_{f2})^2][M^2-(M_{f1}+M_{f2})^2]}/(2M)$
which is the three momentum of the final mesons.

\section{Numerical results and discussions}
In order to fix the parameters in Cornell potential in Eq.(\ref{eq16}) and masses of quarks,
 we take $a=e=2.7183,
\lambda=0.21$ GeV$^2$, ${\Lambda}_{QCD}=0.27$ GeV, $\alpha=0.06$
GeV, $m_c=1.62$ GeV, $m_u=0.305$ GeV, $m_d=0.311$ GeV, $m_s=0.500$ GeV, $etc$~\cite{w1,mass1},
which give the best to fit the mass spectra of the ground charmonium states and other heavy meson states.
The corresponding mass spectra of the $P$-wave charmonium states are shown in Table~\ref{mass} which are obtained by solving the coupled Salpeter
equations Eq.(\ref{eq11}).
\begin{table}\caption{\label{mass}The Mass spectra of $P$-wave charmonia (unit in MeV)~\cite{mass1}.
`Ex.' means the experimental data from PDG~\cite{PDG}.}
\begin{center}
\begin{tabular}{ccccc}
\hline
$(n^{2S+1}L_J)J^{PC}$&Our results&screened potential model\cite{th8}& GI(NR) model\cite{Godfrey}&Ex  \\ \hline
$(1^3P_0)0^{++}$&3414.7(input)&3433&3445(3424)&$3414.71\pm0.30$\\
$(2^3P_0)0^{++}$&3836.8&3842&3916(3852)&--\\
$(3^3P_0)0^{++}$&4140.1&4131&4292(4202)&--\\
$(1^3P_1)1^{++}$&3510.3(input)&3510&3510(3505)&$3510.67\pm0.05$\\
$(2^3P_1)1^{++}$&3928.7&3901&3953(3925)&--\\
$(3^3P_1)1^{++}$&4228.8&4178&4317(4271)&--\\
$(1^3P_2)2^{++}$&3556.1(input)&3554&3550(3556)&$3556.17\pm0.07$\\
$(2^3P_2)2^{++}$&3972.4&3937&3979(3972)&--\\
$(3^3P_2)2^{++}$&4271.0&4208&4337(4317)&--\\
$(1^1P_1)1^{+-}$&3526.0(input)&3519&3517(3516)&$3525.38\pm0.11$\\
$(2^1P_1)1^{+-}$&3943.0&3908&3956(3934)&--\\
$(3^1P_1)1^{+-}$&4242.4&4184&4318(4279)\\
\hline
\end{tabular}
\end{center}
\end{table}

\subsection{$\chi_{c0}(2P)$ and $\chi_{c0}(3P)$}
\begin{table}[htbp]\caption{\label{3p0 result}The strong decay type and decay widths of the $\chi_{c0}(2P)$ and $\chi_{c0}(3P)$ (unit in MeV).
The results in the parentheses are calculated with $M_{\chi_{c0}(2P)}=3862.0$ MeV and $M_{\chi_{c0}(3P)}=4156.0$ MeV.}
\begin{center}
\begin{tabular}{cccccccc}
\hline
Meson&State&Mode &Our Result &LP(SP)\cite{zhaoqiang}&\cite{Godfrey}&\cite{liuxiang2021}&\cite{Swanson}\\ \hline
$\chi_{c0}$&$2^3P_0$&$D\bar D$ &21.0(16.4)&22(28)&30&--&23\\\hline
&$3^3P_0$&$D\bar D$ &0.17(0.13)&0.04(0.08)&0.5&2&--\\
&&$D^*\bar D^*$ &79.0(69.4)&21(30)&43&67&--\\
&&$D_sD_s$ &1.7(1.8)&8.9(9)&6.8&3&--\\
&&$D_s^*D_s^*$&--&2.7(--)&--&--&--\\
&&total&81(71)&33(39)&51&72&--\\
\hline
\end{tabular}
\end{center}
\end{table}

First, we study the higher charmonium states with $J^{PC}=0^{++}$, $\chi_{c0}(2P)$ and $\chi_{c0}(3P)$,
and give the two-body open charm strong decay results in Table~\ref{3p0 result}.
The mass $M_{\chi_{c0}(2P)}=3836.8$ MeV is close to the result of the
screened potential model (3842 MeV)~\cite{th8} and the nonrelativistic potential model (3852 MeV)~\cite{Godfrey}.
Limited by phase space , there is one decay mode $0^{++}\to 0^-0^-$ for the $\chi_{c0}(2P)$,
the corresponding decay channel only includes $D\bar D$ within the kinematic ranges.
And we also get the total strong decay widths of the $\chi_{c0}(2P)$: $\Gamma=21.0$ MeV,
which is in accordance with the result of the linear potential quark model (22 MeV)~\cite{zhaoqiang} and the usual $^3P_0$ model (23 MeV)~\cite{Swanson}.

Belle reported a charmonium-like state $X^*(3860)$,
and they claimed the $X^*(3860)$ seems to be a candidate of
the $\chi_{c0}(2P)$ state~\cite{3860}.
The Ref.~\cite{zhaoqiang} studied the strong decays of the $X^*(3860)$ as $\chi_{c0}(2P)$ by LP and SP models,
and they gave a similar value $\Gamma\approx22\sim28$ MeV.
The analysis of Ref.~\cite{th5} showed that the $X^*(3860)$ was an indication of the $\chi_{c0}(2P)$ state.
Assuming the $X^*(3860)$ as the $\chi_{c0}(2P)$,
the Ref.~\cite{x3860zgwang} calculated the strong decay of the $\chi_{c0}(2P)$,
the total decay of the $X^*(3860)$ state ranged from 110
to 180 MeV with $R=2.3\sim 2.5$ GeV$^{-1}$,
the corresponding decay mode and total decay
width were consistent with the experimental data.
Taking the $X^*(3860)$ as the $\chi_{c0}(2P)$,
we get the strong decay width $\Gamma=16.4$ MeV,
which is smaller than the center value of the $X^*(3860)$: $(201^{+154}_{-67}$$^{+88}_{-82})$ MeV.
But the strong decay width is related to the parameter $\gamma$ in the $^3P_0$ model,
the result increases with the parameter $\gamma$.
In addition,
considering the large uncertainties of the mass and decay width for the $X^*(3860)$,
we can't exclude that the $X^*(3860)$ is $\chi_{c0}(2P)$.
And more investigations are needed to confirm the property of the $X^*(3860)$ in the future.

By solving the Eq.~(\ref{eq11}), we get the mass of $\chi_{c0}(3P)$ as: $M_{\chi_{c0}(3P)}=4140.1$ MeV,
which is close to the result of the screened potential model (4131 MeV)~\cite{th8}.
The dominant strong decay of the is OZI-allowed two-body open charm strong decay.
And there are two decay types: $0^{++}\to 0^-0^-$ and $0^{++}\to 1^-1^-$,
while decay mode $0^{++}\to 0^-1^-$ is forbidden.
Therefore, the final mesons include $D\bar D$, $D_sD_s$ and $D^*\bar D^*$  within the kinematic ranges,
and there is no $D\bar D^*$.
The decay channel $\chi_{c0}(3P)\to D\bar D$ has the largest phase space,
but due to the node structure of $\chi_{c0}(3P)$'s wave functions,
the integrand (which consists of the overlapped wave functions) oscillates accordingly in the amplitude.
The positive contribution of the integrand almost cancels the negative contribution in $\chi_{c0}(3P)\to D\bar D$,
leading to a smallest value of $\chi_{c0}(3P)\to D\bar D$ in Table~\ref{3p0 result}.
So the dominant contribution comes from $\chi_{c0}(3P)\to D^*\bar D^*$,
which is consistent with the result of the screened potential model~\cite{th8}.
Then we calculate all of the two-body open charm strong decays,
the total strong decay width $\Gamma_{\chi_{c0}(3P)}=81$ MeV is close to the result of the unquenched quark model (71 MeV)~\cite{liuxiang2021}.

The $X(4160)$ was observed by Belle from
the process $e^+e^-\to J/\psi D^*\bar D^*$~\cite{39404160}.
The Ref.~\cite{th8} and \cite{zhao1} discussed possible interpretations for the $X(4160)$ based on the NRQCD calculations and the potential model, two likely assignments for the $X(4160)$ were $\chi_{c0}(3P)$ and $\eta_c(4S)$.
The Ref.~\cite{x41601} calculated the strong decays of the $\eta_c(nS)$,
they found that the explanation of the $X(3940)$
as the $\eta_c(3S)$ is possible and the assignment of the $X(4160)$ as the $\eta_c(4S)$ can not
be excluded.
The Ref.~\cite{416011} calculated the strong decay of the $X(4160)$ which was assumed as the $\chi_{c0}(3P)$,
$\chi_{c1}(3P)$, $\eta_{c2}(2D)$ or $\eta_c(4S)$ by the $^3P_0$ model,
they thought that the excited charmonium state $\eta_c(4S)$ cannot be ruled out as
an assignment for the $X(4160)$.
Considering the $X(4160)$ as the $\eta_c(4S)$ state, we also calculated the strong decay of the $\eta_c(4S)$ in Ref.~\cite{4S4160},
the ratio of the decay width $\frac{\Gamma(D\bar D^*)}{\Gamma (D^*\bar D^*)}$ of $\eta_c(4S)$ was larger than
the experimental data of the $X(4160)$,
thus,
the $X(4160)$ was not the candidate of the $\eta_c(4S)$.
In this work, the mass of $\chi_{c0}(3P)$ is close to the $X(4160)$,
assigning the $X(4160)$ as the $\chi_{c0}(3P)$,
the total strong decay width $\Gamma_{\chi_{c0}(3P)}=71$ MeV is rough consistent with the result of experimental
results for $X(4160)$: $(139^{+111}_{-61}\pm21)$ MeV.
Then we calculate relative ratios $\frac{\Gamma(\chi_{c0}(3P)\to D\bar D)}{\Gamma(\chi_{c0}(3P)\to D^*\bar D^*)}\approx0.0019<0.09$,
and $\frac{\Gamma({\chi_{c0}(3P)\to D\bar D^*})}{\Gamma({\chi_{c0}(3P)\to D^*\bar D^*})}=0<0.22$,
both of which agree with the experimental results of $X(4160)$ by Belle collaboration~\cite{39404160}.
Therefore,
$X(4160)$ is a good candidate for $\chi_{c0}(3P)$.

\subsection{$\chi_{c1}(2P)$ and $\chi_{c1}(3P)$}

Using the BS method, we get the masses and wave functions of the $\chi_{c1}(2P)$ and $\chi_{c1}(3P)$.
The $M_{\chi_{c1}(2P)}=3928.7$ MeV is close to the result of the nonrelativistic potential model (3925 MeV)~\cite{Godfrey}.
The two-body open charm strong decay results
have been shown in Table~\ref{3p1 result}.
For the $\chi_{c1}(2P)$, there is only one decay mode $1^{++}\to 0^-1^-$,
so the final state include $DD^*$ state.
The corresponding decay width $\Gamma_{\chi_{c1}(2P)}=103$ MeV is the same as the results of linear potential quark model (102 MeV)~\cite{zhaoqiang},
but smaller than the results of other methods.

The mass $M_{\chi_{c1}(3P)}=4228.8$ MeV is larger than the result of
the screened potential model (4178 MeV)~\cite{th8},
but smaller the than results of the relativized Godfrey-Isgur model (4317 MeV) and nonrelativistic potential model(4271) MeV~\cite{Godfrey}.
The two-body open charm strong decay of $\chi_{c1}(3P)$ have three decay modes: $1^{++}\to 0^-1^-,\;1^-1^-,\;0^-0^+$
and include five final states: $DD^*$, $D^*\bar D^*$, $D_sD_s^*$, $D_s^*D_s^*$, $DD_0$.
The dominant strong decay channels are $DD^*$, $D^*\bar D^*$ and $D_sD_s^*$.
The ratios between different partial width are independent of the strength parameter $\gamma$,
they are $\frac{\Gamma(\chi_{c1}(3P)\to D^*\bar D^*)}{\Gamma(\chi_{c1}(3P)\to DD^*)}\approx0.63$ and
$\frac{\Gamma(\chi_{c1}(3P)\to D_sD_s^*)}{\Gamma(\chi_{c1}(3P)\to DD^*)}\approx0.42$,
which can be explored in the future experiment.
The total strong decay width $\Gamma_{\chi_{c1}(3P)}=29.7$ MeV,
which is similar to the result of the linear potential quark model (23 MeV) and the relativized
Godfrey-Isgur model (39 MeV)~\cite{Godfrey},
but smaller than the result of the unquenched quark model (48 MeV)~\cite{liuxiang2021}.

\begin{table}[htbp]\caption{\label{3p1 result}The strong decay type and decay widths of the $\chi_{c1}(2P)$ and $\chi_{c1}(3P)$ (unit in MeV).
The results in the parentheses are calculated with $M_{\chi_{c1}(3P)}=4274.0$ MeV.}
\begin{center}
\begin{tabular}{ccccccccc}
\hline
Meson&State&Mode &Our Result&\cite{th3} &LP(SP)\cite{zhaoqiang}&\cite{Godfrey}&\cite{liuxiang2021}&\cite{Swanson} \\ \hline
$\chi_{c1}$&$2^3P_1$&$D D^{*}$ &103&--&102(127)&165&--&127\\\hline
&$3^3P_1$&$DD^*$ &14.3(24.9)&6.6&7.1(5.3)&6.8&20&--\\
&&$D^*\bar D^*$ &9.0(14.7)&28&0.2(1.1)&19&26&--\\
&&$D_sD_s^*$ &6.0(10.2)&6.3&11(8.0)&9.7&--&--\\
&&$D_s^*D_s^*$&0.4(2.9)&2.5&5.5(--)&2.7&2&--\\
&&$DD_0$&0.01(0.02)&0.2&0.001(--)&0.1&--&--\\
&&total&29.7(52.7)&43.6&23(14)&39&48&--\\
\hline
\end{tabular}
\end{center}
\end{table}

Some charmonium-like states with $J^{PC}=1^{++}$: $X(3872)$,
$X(4274)$, have been discovered in experiments~\cite{PDG}.
They may be the good candidates for the $\chi_{c1}(nP)$.
The Ref.~\cite{3872-1} calculated the E1 radiative and strong decays of the $X(3872)$ as all possible $1D$ and $2P$
$c\bar c$ states.
The Ref.~\cite{3872-2} explored the $1^3D_2$, $1^3D_3$ and $2^1P_1$ charmonium candidates for $X(3872)$, and the $1^3D_2$, $1^3D_3$ were favored candidates for the $X(3872)$, both have prominent radiative decays.
The $X(3872)$ was examined by the molecule model and the charmonium model in Ref.~\cite{3872-3},
the author thought that the $X(3872)$ may fit more
likely to the excited $^3P_1$ charmonium than to the molecule.
The quantum number of the $X(3872)$ is the same as $\chi_{c1}(2P)$,
but its mass is about 50 MeV lighter than the result of our prediction.
Considering the $X(3872)$ as the $\chi_{c1}(2P)$,
we have calculated the radiative $E1$ decay widths of the $X(3872)$ through the BS method in Ref.~\cite{thwang1},
the result was in agreement with the experimental data.
However, there is no two-body open charm strong decay for $X(3872)$ which has a narrow width $\Gamma_{X(3872)}<1.2$ MeV~\cite{PDG}.
Because the mass of $X(3872)$ happens to lie around the $D\bar D^*$ threshold,
many authors believed that it's an ideal candidate for the $D\bar D^*$
exotic hadrons.
Various scenarios have been discussed in the literature Ref.~\cite{Olsen,fkguo,swanson1,Ferretti,slzhu2},
but until now, the nature of the $X(3872)$ still remains unclear,
many more investigations are very essential to understand the property of the $X(3872)$ in the future.

The Ref.~\cite{th3} calculated the strong decay of $X(4274)$ as $\chi_{c1}(3P)$ with $M=4317$ MeV, and they got the width $\Gamma=43.6$ MeV.
The Ref.~\cite{liuxiang2021} gave the decay width of $\chi_{c1}(3P)$: $\Gamma=48$ MeV,
which was consistent with experimental data,
and it was possible to assign $X(4274)$ as the $\chi_{c1}(3P)$ state.
Considering the $X(4274)$ as the $\chi_{c1}(3P)$ state,
we calculate its strong decays by the BS method.
The total strong decay width $\Gamma=52.7$ MeV is larger than 29.7 MeV with $M_{\chi_{c1}(3P)}=4228.8$ MeV,
so the strong decay width is sensitive to the mass of the $\chi_{c1}(3P)$.
The total strong decay width $\Gamma=52.7$ MeV is agreement with
the world average data $\Gamma_{X(4274)}=(49\pm12)$ MeV~\cite{PDG}.
Therefore,
the $X(4274)$ is a good candidate for the $\chi_{c1}(3P)$ state.
Assigning the $X(4274)$ as the $\chi_{c1}(3P)$,
the relative ratios are
$\frac{\Gamma(\chi_{c1}(3P)\to D^*\bar D^*)}{\Gamma(\chi_{c1}(3P)\to DD^*)}\approx0.59$ and
$\frac{\Gamma(\chi_{c1}(3P)\to D_sD_s^*)}{\Gamma(\chi_{c1}(3P)\to DD^*)}\approx0.41$,
which can provide more useful information to observe the $X(4274)$ in the future experiment.

\begin{table}[htbp]\caption{\label{3p2 result}The strong decay type and decay widths of the $\chi_{c2}(2P)$ and $\chi_{c2}(3P)$ (unit in MeV).
The results in the parentheses are calculated with $M_{\chi_{c2}(2P)}=3930.0$ MeV and $M_{\chi_{c2}(3P)}=4350.0$ MeV.}
\begin{center}
\begin{tabular}{cccccccc}
\hline
Meson&State&Mode &Our Result &LP(SP)\cite{zhaoqiang}&\cite{Godfrey}&\cite{liuxiang2021}&\cite{Swanson} \\ \hline
$\chi_{c2}$&$2^3P_2$&$D\bar D$ &24.6(20.4)&--&42&--&32\\
&&$D D^{*}$ &21.1(6.9)&--&37&--&28\\
&&$D_sD_s$ &0.6(--)&--&0.7&--&0.5\\
&&total &46(27.3)&--&80&--&60.5\\  \hline
&$3^3P_2$&$D\bar D$ &5.7(8.7)&8.1(7.3)&8.0&7&--\\
&&$DD^*$ &2.2(2.7)&17(13)&2.4&3&--\\
&&$D^*\bar D^*$ &24.3(28.6)&4.2(7.1)&24&39&--\\
&&$D_sD_s$ &0.4(0.9)&1.0(0.6)&0.8&0&--\\
&&$D_sD_s^*$ &3.0(4.2)&0.3(1.4)&11&--&--\\
&&$D_s^*D_s^*$&0.6(1.8)&4.8(--)&7.2&1&--\\
&&$DD_1$&--(0.9)&1.0(0)&1.1&--&--\\
&&$DD_1^{\prime}$&--(10.9)&7.3(0)&12&--&--\\
&&$DD_2$&--(1.4)&--&--&--&--\\
&&$D^*D_0$&--(0.3)&--&--&--&--\\
&&total&36(60)&43(30)&66&50&--\\
\hline
\end{tabular}
\end{center}
\end{table}

\subsection{$\chi_{c2}(2P)$ and $\chi_{c2}(3P)$}

By solving the Eq.(\ref{eq11}),
we also obtain the mass of $\chi_{c2}(2P)$ and $\chi_{c2}(3P)$,
the $M_{\chi_{c2}(2P)}=3972.4$ MeV is in accordance with the
relativized Godfrey-Isgur model (3979 MeV) and the nonrelativistic potential model (3972 MeV)~\cite{Godfrey}.
The $M_{\chi_{c2}(3P)}=4271.0$ MeV is smaller than the mass in the
relativized Godfrey-Isgur model (4337 MeV) and the nonrelativistic potential model (4317 MeV)~\cite{Godfrey}.
The two-body open charm strong decays of the $\chi_{c2}(2P)$ and $\chi_{c2}(3P)$,
are shown in Table~\ref{3p2 result}.
The $\chi_{c2}(2P)$ state has two decay modes: $2^{++}\to0^-0^-,\;0^-1^-$,
and the corresponding final states include $D\bar D$, $D D^{*}$ and $D_sD_s$.
We find that the dominant channels are $D\bar D$ and $D D^{*}$,
the $\chi_{c2}(2P)\to D_sD_s$ is very small
with the small phase space.
Then the total strong decay width $\Gamma_{\chi_{c2}(2P)}=46$ MeV,
which is smaller than the results of the nonrelativistic potential model (80 MeV)~\cite{Godfrey} and the usual $^3P_0$ model (60.5 MeV)~\cite{Swanson}.

There are three decay modes for the $\chi_{c2}(3P)$: $2^{++}\to0^-0^-,\;0^-1^-,\;1^-1^-$,
and six strong decay channels $D \bar D$, $D D^{*}$,
$D^*\bar D^*$,$D_sD_s$, $D_sD_s^*$ and $D_s^*D_s^*$.
The total strong decay width $\Gamma=36$  MeV is consistent with the results of
linear potential quark model (43 MeV) and screened potential quark model (30 MeV)~\cite{zhaoqiang}.
$\chi_{c2}(3P)\to D^*\bar D^*$ is the dominant decay channel,
which contribute about $68\%$ of the total strong decay width.
We also predict the relative ratios $\frac{\Gamma(\chi_{c2}(3P)\to D\bar D)}{\Gamma(\chi_{c2}(3P)\to D^*\bar D^*)}\approx0.23$
and $\frac{\Gamma(\chi_{c2}(3P)\to DD^*)}{\Gamma(\chi_{c2}(3P)\to D^*\bar D^*)}\approx0.091$ with $M_{\chi_{c2}(3P)}=4271.0$ MeV.

The Ref.~\cite{th11} studied the strong decays of the $\chi_{cJ}(2P)$ and $\chi_{cJ}(3P)$,
the mass of $\chi_{c2}(2P)$ was very close to the experimental data of the $Z(3930)$,
but the decay width $\Gamma\approx 68$ MeV,
which was 3 times that of the experimental value.
The $Z(3930)$ was assigned to the $\chi_{c2}(2P)$,
and the strong decay width was $\Gamma=19.0$ MeV in the Ref.~\cite{39301}.
Assigning the $Z(3930)$ as the $\chi_{c2}(2P)$,
we have taken two methods to calculate the OZI-allowed two-body strong
decay processes of the $\chi_{c2}(2P)$ state in detail: the BS method and the extended $^3P_0$ model in Ref.~\cite{thwang}.
The total decay width is consistent with the experimental data,
which means the $Z(3930)$ is a good candidate for the $\chi_{c2}(2P)$,
so we only list the total strong decay width of the $\chi_{c2}(2P)$ in this paper.

The $X(4350)$ was observed by Belle in the $\phi J/\psi$ mass spectrum,
which is a candidate for the $\chi_{c2}(3P)$~\cite{4350}.
The open-charm decay of $\chi_{c2}(3P)$ with $R=1.9\approx2.3$ GeV$^{-1}$
was well consistent with experimental data of the $X(4350)$,
which showed that the $X(4350)$
as a good candidate of $\chi_{c2}(3P)$ in Ref.~\cite{th13}.
Assigning
the $X(4350)$ as the $\chi_{c2}(3P)$ state,
the Ref.~\cite{zhaoqiang} studied its the strong decay,
and got the decay width which was about 90 MeV.
Considering the $X(4350)$ as the $\chi_{c2}(3P)$,
some new decay channels are allowed with the increasing mass,
such as $DD_1$, $DD_1^\prime$, $DD_2$ and $D^*D_0$.
So the total decay width increases to $\Gamma=60$ MeV,
which is consistent with the results of the nonrelativistic potential model (66 MeV)~\cite{Godfrey},
but larger than the result of experimental dada $\Gamma_{X(4350)}=(13^{+18}_{-9}\pm4)$ MeV.
Thus, if one takes the $\chi_{c2}(3P)$ as an assignment of the $X(4350)$,
the precision measurements are needed in further experiments.
The relative ratios $\frac{\Gamma(\chi_{c2}(3P)\to D\bar D)}{\Gamma(\chi_{c2}(3P)\to D^*\bar D^*)}\approx0.30$
and $\frac{\Gamma(\chi_{c2}(3P)\to DD^*)}{\Gamma(\chi_{c2}(3P)\to D^*\bar D^*)}\approx0.094$ with $M_{\chi_{c2}(3P)}=4350.0$ MeV,
also can provide evidence to discover the $X(4350)$ for the future experiment.

\begin{table}[htbp]\caption{\label{1p1 result}The strong decay type and decay widths of the $h_{c}(2P)$ and $h_{c}(3P)$ (unit in MeV).}
\begin{center}
\begin{tabular}{ccccccc}
\hline
Meson&State&Mode &Our Result &LP(SP)\cite{zhaoqiang}&\cite{Godfrey}&\cite{Swanson} \\ \hline
$h_c$&$2^1P_1$&$DD^*$ &48&64(68)&87&67\\\hline
&$3^1P_1$&$DD^*$ &16.0&14(11)&3.0&--\\
&&$D^* \bar D^*$ &3.6&4.8(7.5)&22&--\\
&&$D_sD_s^*$ &6.4&6.5(6.3)&15&--\\
&&$D_s^*D_s^*$&0.2&3.6(--)&7.5&--\\
&&$DD_0$&4.6&15(5.0)&28&--\\
&&total&31&44(30)&75&--\\
\hline
\end{tabular}
\end{center}
\end{table}

\subsection{$h_{c}(2P)$ and $h_{c}(3P)$}

Finally, we study the higher charmonium states with $J^{PC}=1^{+-}$, $h_c(2P)$ and $h_c(3P)$.
The $M_{h_c(2P)}=3943.0$ MeV is consistent with the mass of
the relativized Godfrey-Isgur model (3956 MeV) and the nonrelativistic potential model (3934 MeV)~\cite{Godfrey}.
The $M_{h_c(3P)}=4242.4$ MeV is close to the results of the nonrelativistic potential model (4279 MeV)~\cite{Godfrey}.
And the two-body open charm strong decay results
are shown in Table~\ref{1p1 result}.
The $h_c(2P)$ has one decay mode $1^{+-}\to 0^-1^-$, and  only can decay into $DD^*$.
The total strong decay width $\Gamma_{h_{c}(2P)}=48$ MeV,
which is smaller than the results of other theoretical models.

The main decay modes of the $h_c(3P)$ include $1^{+-}\to 0^-1^-,\;1^-1^-,\;0^-0^+$.
The final states $DD^*$, $D^*\bar D^*$, $D_sD_s^*$, $D_s^*D_s^*$, $DD_0$ are allowed.
The total strong decay width $\Gamma_{h_{c}(3P)}=31$ MeV is in accordance with the result of the screened potential quark model (30 MeV)~\cite{zhaoqiang}.
The main decay channel $h_c(3P)\to DD^*$ contributes about $52\%$ of the total strong decay width.
The corresponding relative ratios: $\frac{\Gamma(h_{c}(3P)\to D^*\bar D^*)}{\Gamma(h_{c}(3P)\to DD^*)}\approx0.23$
and $\frac{\Gamma(h_{c}(3P)\to D_sD_s^*)}{\Gamma(h_{c}(3P)\to DD^*)}\approx0.40$
can provide theoretical
assistance to confirm the $h_c(3P)$ in future experiments.

\section{summary}
In conclusion, we have studied the two-body open charm strong decays of the
$2P$ and $3P$ charmonium states by the BS method combined with the $^3P_0$ model.
The wave functions and mass spectra of the initial $2P$ and $3P$ charmonium states are obtained
by solving the BS equation with the relativistic correction.
Considering the relativistic correction,
the masses of some $3P$ charmonium states in our model will have a difference with the results of the nonrelativistic potential model.
Then we get the two-body open charm strong decay widths of the $2P$ and $3P$ charmonium states.

Considering the $X^*(3860)$ as $\chi_{c0}(2P)$,
the narrow strong decay width is smaller than the experimental data,
because the strong decay width depend on the parameter,
and there are large errors in the mass and the width of the $X^*(3860)$,
we cannot rule out that the $X^*(3860)$ is $\chi_{c0}(2P)$.
Taking the $X(4160)$ as the $\chi_{c0}(3P)$,
the total strong decay width is in accordance with the experimental
result of the $X(4160)$ with the uncertainty.
In addition, because of the node structure of the $\chi_{c0}(3P)$'s wave functions,
the integrand oscillates accordingly in the amplitude,
the decay $X(4160)\to D\bar D$ is strong suppressed.
Then the relative ratios $\frac{\Gamma(\chi_{c0}(3P)\to D\bar D)}{\Gamma(\chi_{c0}(3P)\to D^*\bar D^*)}\approx0.0019<0.09$,
and $\frac{\Gamma({\chi_{c0}(3P)\to D\bar D^*})}{\Gamma({\chi_{c0}(3P)\to D^*\bar D^*})}=0<0.22$ are consistent with the experimental results.
Therefore, the $X(4160)$ is a good candidate for the $\chi_{c0}(3P)$.

Assigning the $X(4274)$ as the $\chi_{c1}(3P)$,
we find that the total strong decay width is in agreement with the data of the $X(4274)$,
so the $X(4274)$ resonance is a good candidate for the $\chi_{c1}(3P)$.
The relative ratios $\frac{\Gamma(\chi_{c1}(3P)\to D^*\bar D^*)}{\Gamma(\chi_{c1}(3P)\to DD^*)}\approx0.59$ and
$\frac{\Gamma(\chi_{c1}(3P)\to D_sD_s^*)}{\Gamma(\chi_{c1}(3P)\to DD^*)}\approx0.41$ can provide
useful information in experiments.

The $Z(3930)$ has been confirmed as the $\chi_{c2}(2P)$ state in our previous work,
so we only show the strong decay width in this work.
Considering the $X(4350)$ as $\chi_{c2}(3P)$,
we find that the total strong decay width of the $\chi_{c2}(3P)$ is larger than the result of the $X(4350)$,
thus, if we take the $\chi_{c2}(3P)$ as an assignment of the $X(4350)$,
we need many more investigations in the future.

Finally,
we also calculate the two-body open charm strong decay of the $h_{c}(2P)$ and $h_{c}(3P)$,
and give the total strong decay widths of the $h_{c}(2P)$ and $h_{c}(3P)$.
The corresponding relative ratios $\frac{\Gamma(h_{c}(3P)\to D^*\bar D^*)}{\Gamma(h_{c}(3P)\to DD^*)}\approx0.23$
and $\frac{\Gamma(h_{c}(3P)\to D_sD_s^*)}{\Gamma(h_{c}(3P)\to DD^*)}\approx0.40$ can
give the theoretical
assistance in future experiments.

 \noindent
{\Large \bf Acknowledgements}
We would like to thank Shi-Lin Zhu for many valuable
discussions and assistance during this work.
This work was supported by
the National Natural Science Foundation of China (NSFC) under
Grant No.~11865001 and No.~12075073,
and the CAS "Light of West China" Program.

\appendix{\section{Instantaneous Bethe-Salpeter Equation}

The BS equation which is used to describe the heavy mesons can be written as~\cite{BS}:
\begin{equation}
(\not\!{p_{1}}-m_{1})\chi(q)(\not\!{p_{2}}+m_{2})=
i\int\frac{d^{4}k}{(2\pi)^{4}}V(P,k,q)\chi(k)\;, \label{eq1}
\end{equation}
where $\chi(q)$ and $P$ are the wave function and the momentum of the bound state, respectively.
$q$ is the relative momentum between quark and anti-quark in meson,
$p_{1}=\frac{m_1}{m_1+m_2}P+q$,
$p_{2}=\frac{m_2}{m_1+m_2}P-q$ and $m_1$, $m_2$ are the momentum and the mass of the quark and anti-quark, respectively.
The $V(P,k,q)$ is the
interaction kernel between the quark and antiquark.

In order to solve the Eq.~(\ref{eq1}),
the instantaneous approximation is adopted in the interaction kernel $V(P,k,q)$~\cite{Salp}:
$$V(P,k,q) \Rightarrow V(|\vec k-\vec q|)\;.$$

For convenience, the relative momentum $q$ is decomposed into two parts
$q_{\parallel}$ and $q_{\perp}$,
$$q^{\mu}=q^{\mu}_{\parallel}+q^{\mu}_{\perp}\;,$$
$$q^{\mu}_{\parallel}\equiv (q_P/M)P^{\mu}\;,\;\;\;
q^{\mu}_{\perp}\equiv q^{\mu}-q^{\mu}_{\parallel}\;,\;\;\;q_P=P\cdot q/M.$$

Then the Eq.~(\ref{eq1}) can be expressed as:
\begin{equation}
\chi(q_{\parallel},q_{\perp})=S_{1}(p_{1})\eta(q_{\perp})S_{2}(p_{2})\;.
\label{eq6}
\end{equation}
$\eta(q^{\mu}_{\perp})$ is related to three dimensional BS wave function $\varphi_{p}(q^{\mu}_{\perp})$ as
follows:
$$\varphi_{P}(q^{\mu}_{\perp})\equiv i\int
\frac{{\rm d}q_{p}}{2\pi}\chi(q^{\mu}_{\parallel},q^{\mu}_{\perp})\;,$$
\begin{eqnarray}
\eta(q^{\mu}_{\perp})\equiv\int\frac{{\rm d}k_{\perp}}{(2\pi)^{3}}
V(k_{\perp},q_{\perp})\varphi_{p}(k^{\mu}_{\perp})\;,
 \label{eq5}
\end{eqnarray}
$S_{1}(p_{1})$ and $S_{2}(p_{2})$ are the propagators of the quark and anti-quark which can be decomposed as:
\begin{equation}
S_{i}(p_{i})=\frac{\Lambda^{+}_{ip}(q_{\perp})}{J(i)q_{p}
+\alpha_{i}M-\omega_{i}+i\epsilon}+
\frac{\Lambda^{-}_{ip}(q_{\perp})}{J(i)q_{p}+\alpha_{i}M+\omega_{i}-i\epsilon}\;,
\label{eq7}
\end{equation}
with
\begin{equation}
\omega_{i}=\sqrt{m_{i}^{2}+q^{2}_{_T}}\;,\;\;\;
\Lambda^{\pm}_{ip}(q_{\perp})= \frac{1}{2\omega_{ip}}\left[
\frac{\not\!{P}}{M}\omega_{i}\pm
J(i)(m_{i}+{\not\!q}_{\perp})\right]\;, \label{eq8}
\end{equation}
where $i=1, 2$ for quark and anti-quark, respectively,
 and
$J(i)=(-1)^{i+1}$.

The positive and negative energy projected wave functions $\varphi^{\pm\pm}_{p}(q_{\perp})$ are defined as:
\begin{equation}
\varphi^{\pm\pm}_{p}(q_{\perp})\equiv
\Lambda^{\pm}_{1p}(q_{\perp})
\frac{\not\!{P}}{M}\varphi_{p}(q_{\perp}) \frac{\not\!{P}}{M}
\Lambda^{{\pm}}_{2p}(q_{\perp})\;. \label{eq10}
\end{equation}

Then under instantaneous approximation,
with contour integration over $q_{p}$ on both sides of
Eq.~(\ref{eq6}), we obtain the full Salpeter equation:
$$
(M-\omega_{1}-\omega_{2})\varphi^{++}_{p}(q_{\perp})=
\Lambda^{+}_{1p}(q_{\perp})\eta_{p}(q_{\perp})\Lambda^{+}_{2p}(q_{\perp})\;,
$$
$$(M+\omega_{1}+\omega_{2})\varphi^{--}_{p}(q_{\perp})=-
\Lambda^{-}_{1p}(q_{\perp})\eta_{p}(q_{\perp})\Lambda^{-}_{2p}(q_{\perp})\;,$$
\begin{equation}
\varphi^{+-}_{p}(q_{\perp})=\varphi^{-+}_{p}(q_{\perp})=0\;.
\label{eq11}
\end{equation}

The wave functions are different for the bound states with different quantum $J^{PC}$ (or $J^{P}$).
First, we give the original BS wave functions for the different bound state, then reduce the wave functions through the last
equation of Eq.~(\ref{eq11}). Finally the numerical result of the wave functions and mass spectrum are obtained
by solving the first and second equations in Eq.~(\ref{eq11}).
And the detailed solution of the Salpeter equation also has been discussed in Ref.~\cite{w1,BS1,glwang,mass1}.

To solve the Eq.~(\ref{eq11}), we take the Cornell
potential as the instantaneous interaction kernel $V$,
which include a linear scalar interaction and a vector interaction.
In the momentum space and the C.M.S of the bound state,
the interaction potential is read as:
$$V(\vec q)=V_s(\vec q)
+\gamma_{_0}\otimes\gamma^0 V_v(\vec q)~,$$
\begin{equation}
V_s(\vec q)=-(\frac{\lambda}{\alpha}+V_0) \delta^3(\vec
q)+\frac{\lambda}{\pi^2} \frac{1}{{(\vec q}^2+{\alpha}^2)^2}~,
~~V_v(\vec q)=-\frac{2}{3{\pi}^2}\frac{\alpha_s( \vec q)}{{(\vec
q}^2+{\alpha}^2)}~,\label{eq16}
\end{equation}
where $\lambda$ is the string constant and $\alpha_s(\vec q)=\frac{12\pi}{33-2N_f}\frac{1}
{\log (a+{\vec q}^2/\Lambda^{2}_{QCD})}$ is the running coupling constant. In order to fit the data of
heavy quarkonia, a constant $V_0$ is often added to confining
potential. We also introduce a small
parameter $a$ to
avoid the divergence in the denominator. The constants $\lambda$, $\alpha$, $V_0$ and
$\Lambda_{QCD}$ are the parameters that characterize the potential.

\section{The relativistic wave functions}

In this paper, we focus on the OZI-allowed two-body open charm strong decay of the $2P$ and $3P$ charmonium states.
The detailed wave functions have been obtained in Refs.~\cite{w1,mass1,BS1,glwang}.
We mainly introduce the relativistic BS wave functions of the $\chi_{c0}$,
$\chi_{c1}$, $\chi_{c2}$ and $h_c$ in this section.

\subsection{The relativistic BS wave function of the $\chi_{c0}$ with $J^{PC}=0^{++}$}
The original BS wave functions of the $\chi_{c0}$ with $J^{PC}=0^{++}$ can be written as
\begin{equation}\label{chic0}
\varphi_{0^{++}}(q_\perp)=M\left[\frac{\not\!{q}_\perp}{M}f_1(q_\perp)+\frac{\not\!{P}\not\!{q}_\perp}{M^2}f_2(q_\perp)+f_3(q_\perp)+\frac{\not\!{P}}{M}f_4(q_\perp)\right]
\end{equation}
where $M$ is the mass of bound state $\chi_{c0}$,
$f_i(q_\perp)$ is the original radial wave functions that are related to $|\vec q|^2$.
Taking Eq.~(\ref{chic0}) to Eq.~(\ref{eq11}),
the relativistic BS wave functions and the mass
spectrum can be obtained by solving the Salpeter equations Eq.~(\ref{eq11}).
Then the relativistic positive BS wave function is shown as,
\begin{equation}
\varphi^{++}_{0^{++}}(q_\perp) = A_1(q_\perp)+\frac{\not\!{P}}{M}A_2(q_\perp)
+\frac{\not\!{q}_\perp}{M}A_3(q_\perp)+\frac{\not\!{P}\not\!{q}_\perp}{M^2}A_4(q_\perp).
\end{equation}

The corresponding coefficients are
$$A_1 =\frac{(\omega_1+\omega_2)q_\perp^2}{2(m_1\omega_2+m_2\omega_1)}\left[f_1+\frac{m_1+m_2}{\omega_1+\omega_2}f_2\right],\;\;\;
A_2=\frac{(m_1-m_2)q_\perp^2}{2(m_1\omega_2+m_2\omega_1)}\left[f_1+\frac{m_1+m_2}{\omega_1+\omega_2}f_2\right],\;\;$$
$$A_3=\frac{M}{2}\left[f_1+\frac{m_1+m_2}{\omega_1+\omega_2}f_2\right],\;\;\;
A_4=\frac{M}{2}\left[\frac{\omega_1+\omega_2}{m_1+m_2}f_1+f_2\right].$$
where $m_1, m_2$ and
$\omega_1=\sqrt{m_1^{2}+\vec{q}^2},\omega_2=\sqrt{m_2^{2}+\vec{q}^2}$ are
the masses and the energies of the
quark and anti-quark in the $\chi_{c0}$ state.

\subsection{The relativistic BS wave function of the $\chi_{c1}$ with $J^{PC}=1^{++}$}
The original BS wave functions of the $\chi_{c1}$ with $J^{PC}=1^{++}$ is constructed by $P$, ${q}_\perp$
and the polarization vector $\epsilon$,
\begin{equation}\label{chic1}
\varphi_{1^{++}}(q_\perp)=i\varepsilon_{\mu\nu\alpha\beta}\frac{P^\nu}{M} q_\perp^\alpha
\epsilon^\beta\left[g_1\gamma^\mu+g_2\frac{\not\!{P}}{M}\gamma^\mu
+g_3\frac{\not\!{q}_\perp}{M}\gamma^\mu+g_4\frac{\not\!{P}\gamma^\mu\not\!{q}_\perp}{M^2}\right]
\end{equation}
where $\epsilon$ is the polarization vector of the axial vector meson.
The corresponding relativistic positive BS wave function is obtained in Eq.~(\ref{chic1eq}) by solving the Eq.~(\ref{eq11}),
\begin{equation}\label{chic1eq}
\varphi^{++}_{1^{++}}(q_{\perp})=i\varepsilon_{\mu\nu\alpha\beta}\frac{P^\nu}{M} q_\perp^\alpha
\epsilon^\beta\gamma^\mu\left[B_1+B_2\frac{\not\!{P}}{M}
+B_3\frac{\not\!{q}_\perp}{M}+B_4\frac{\not\!{P}\not\!{q}_\perp}{M^2}\right]
\end{equation}
$$B_1 = \frac{1}{2}\left[g_1+\frac{\omega_1+\omega_2}{m_1+m_2}g_2\right],\;\;\;
B_2=-\frac{1}{2}\left[\frac{m_1+m_2}{\omega_1+\omega_2}g_1+g_2\right],$$
$$B_3=\frac{M(\omega_1-\omega_2)}{m_1\omega_2+m_2\omega_1}B_1,\;\;\;
B_4=-\frac{M(m_1+m_2)}{m_1\omega_2+m_2\omega_1}B_1.$$

\subsection{The Relativistic BS Wave function of the $\chi_{c2}$ with $J^{PC}=2^{++}$}

The original BS wave function of the $\chi_{c2}$ is constructed by $P$, ${q}_\perp$,
the polarization tensor $\epsilon_{\mu\nu}$ and the gamma matrices,
\begin{eqnarray}
\varphi_{2^{++}}(q_\perp)&=&\epsilon_{\mu\nu}q_\perp^\mu
q_\perp^\nu\left[h_1(q_\perp)+\frac{\not\!{P}}{M}h_2(q_\perp)
+\frac{\not\!{q}_\perp}{M}h_3(q_\perp)+\frac{\not\!{P}\not\!{q}_\perp}{M^2}h_4(q_\perp)\right]\\ \nonumber
&+&M\epsilon_{\mu\nu}\gamma^\mu q_\perp^\nu\left[h_5(q_\perp)+\frac{\not\!{P}}{M}h_{6}(q_\perp)
+\frac{\not\!{q}_\perp}{M}h_{7}(q_\perp)+\frac{\not\!{P}\not\!{q}_\perp}{M^2}h_{8}(q_\perp)\right]
\end{eqnarray}
where $\epsilon_{\mu\nu}$ is the polarization tensor of the $\chi_{c2}$ with $J^{PC}=2^{++}$.
According to the solve the Eq.~(\ref{eq11}),
we get the relativistic positive BS wave function,
\begin{eqnarray}
\varphi_{2^+}(q_\perp)&=&\epsilon_{\mu\nu}q_\perp^\mu
q_\perp^\nu\left[C_1(q_\perp)+\frac{\not\!{P}}{M}C_2(q_\perp)
+\frac{\not\!{q}_\perp}{M}C_3(q_\perp)+\frac{\not\!{P}\not\!{q}_\perp}{M^2}C_4(q_\perp)\right]\\ \nonumber
&+&M\epsilon_{\mu\nu}\gamma^\mu q_\perp^\nu\left[C_5(q_\perp)+\frac{\not\!{P}}{M}C_{6}(q_\perp)
+\frac{\not\!{q}_\perp}{M}C_{7}(q_\perp)+\frac{\not\!{P}\not\!{q}_\perp}{M^2}C_{8}(q_\perp)\right]
\end{eqnarray}
where the coefficients are
$$C_1 = \frac{1}{2M(m_1\omega_2 + m_2\omega_1)}\left[(\omega_1+\omega_2) q_\perp^2h_3 + (m_1 + m_2) q_\perp^2h_4 + 2M^2\omega_2h_5 -2M^2m_2h_6\right],$$
$$C_2 = \frac{1}{2M(m_1\omega_2 + m_2\omega_1)}\left[(m_1-m_2) q_\perp^2h_3 + (\omega_1 - \omega_2)q_\perp^2h_4 - 2M^2m_2h_5 +2M^2\omega_2f_6\right],$$
$$C_3 = \frac{1}{2}\left[h_3 + \frac{m_1 + m_2}{\omega_1+\omega_2}h_4 -\frac{2M^2}{m_1\omega_2  + m_2\omega_1}h_6\right],
C_4 = \frac{1}{2}\left[\frac{\omega_1+\omega_2}{m_1+m_2}h_3 + h_4 -\frac{2M^2}{m_1\omega_2+m_2\omega_1}h_5\right],$$
$$C_5 = \frac{1}{2}\left[h_5 -\frac{\omega_1+\omega_2}{m_1+m_2}h_6\right],~~~~~~~~~~~~C_6 = \frac{1}{2}\left[-\frac{m_1+m_2}{\omega_1+\omega_2}h_5 +h_6\right],$$
$$C_7 = \frac{M}{2}\frac{\omega_1-\omega_2}{m_1\omega_2+m_2\omega_1}\left[h_5 - \frac{\omega_1+\omega_2}{m_1+m_2}h_6\right],
C_8 = \frac{M}{2}\frac{m_1+m_2}{m_1\omega_2 + m_2\omega_1}\left[-h_5 +\frac{\omega_1+\omega_2}{m_1+m_2}h_6\right]$$

\subsection{The Relativistic BS Wave function of the $h_{c}$ with $J^{PC}=1^{+-}$}
The original BS wave functions of the $h_{c}$ with $J^{PC}=1^{+-}$ also is constructed by $P$, ${q}_\perp$
and the polarization vector $\epsilon$,
\begin{eqnarray}
\varphi_{1^{+-}}(q_\perp)=q_\perp\cdot\epsilon\left[t_1(q_\perp)+\frac{\not\!{P}}{M}t_2(q_\perp)
+\frac{\not\!{q}_\perp}{M}t_3(q_\perp)+\frac{\not\!{P}\not\!{q}_\perp}{M^2}t_4(q_\perp)\right]\gamma_5\
\end{eqnarray}
where $\epsilon$ is the polarization vector of the $h_{c}$.
The corresponding relativistic positive BS wave function is obtained in Eq.~(\ref{chic1eq}) by solving the Eq.~(\ref{eq11}),
\begin{eqnarray}
\varphi^{++}_{1^{+-}}(q_\perp)= q_\perp\cdot\epsilon\left[K_1(q_\perp)+\frac{\not\!{P}}{M}K_2(q_\perp)
+\frac{\not\!{q}_\perp}{M}K_3(q_\perp)+\frac{\not\!{P}\not\!{q}_\perp}{M^2}K_4(q_\perp)\right]\gamma_5
\end{eqnarray}
where the coefficients are
$$K_1 = \frac{1}{2}\left[t_1+\frac{\omega_1+\omega_2}{m_1+m_2}t_2\right],\;\;\;
K_2=\frac{1}{2}\left[\frac{m_1+m_2}{\omega_1+\omega_2}t_1+t_2\right],$$
$$K_3=-\frac{M(\omega_1-\omega_2)}{m_1\omega_2+m_2\omega_1}K_1,\;\;\;
K_4=-\frac{M(m_1+m_2)}{m_1\omega_2+m_2\omega_1}K_1.$$
}

\end{document}